\documentclass{emulateapj}
\usepackage{amsmath}
\usepackage{amssymb}
\usepackage{graphicx}
\usepackage{color}
\usepackage{verbatim}
\usepackage{natbib}
\usepackage{epsfig}

\shorttitle{Calorimetry of GRB 030329}
\shortauthors{Mesler and Pihlstr\"om}

\begin{document}
\title{Calorimetry of GRB 030329: Simultaneous Model Fitting to the Broadband Radio Afterglow and the Observed Image Expansion Rate}


\author{Robert A. Mesler}
\affil{Department of Physics and Astronomy, University of New Mexico MSC07 4220, Albuquerque, NM 87131}
\email{meslerra@unm.edu}

\and

\author{Ylva M. Pihlstr\"om\altaffilmark{1}}
\affil{Department of Physics and Astronomy, University of New Mexico, MSC07 4220, Albuquerque, NM 87131}
\altaffiltext{1}{Ylva Pihlstr\"om is also an Adjunct Astronomer at the National Radio Astronomy Observatory} 




\doublespace

\begin{abstract}
We perform calorimetry on the bright gamma ray burst (GRB) 030329 by fitting simultaneously the broadband radio afterglow and the observed afterglow image size to a semi-analytic magnetohydrodynamical (MHD) and afterglow emission model.  Our semi-analytic method is valid in both the relativistic and non-relativistic regimes, and incorporates a model of the interstellar scintillation that substantially effects the broadband afterglow below 10 GHz.  The model is fitted to archival measurements of the afterglow flux from 1 day to 8.3 years after the burst.  Values for the initial burst parameters are determined and the nature of the circumburst medium is explored.  Additionally, direct measurements of the lateral expansion rate of the radio afterglow image size allow us to estimate the initial Lorentz factor of the jet.
\end{abstract}


\keywords{gamma rays: bursts}


\section{Introduction}

A Gamma ray burst (GRB) afterglow is produced by the interaction of the GRB jet with the circumburst medium in which it is immersed.  An afterglow spectrum is composed of a series of power laws separated by breaks \citep{sariEA98}, and is produced when electrons in the circumburst medium spiral around the tangled and compressed magnetic field lines present at the shock boundary between the jet and the circumburst medium. In this manner, the kinetic energy of the jet is gradually converted into radiation and particle energy \citep{meszarosRees93, katz94}.  The properties of the jet- including its initial kinetic energy, mass, and half-opening angle- only partially determine the temporal evolution of the afterglow.  Because the afterglow luminosity is dependent upon the density of the circumburst medium, the nature of that medium (i.e., whether it is wind-like or a uniform density ISM) plays a crucial role in determining the temporal evolution as well \citep{sariEA98, huangEA00, chevalierLi00, granotSari02, meslerEA12a}.  Analytical models of relativistic jets in power law mediums ($n(r) \propto r^{-k}$, with $k = 0$ for a uniform density and $k = 2$ for a wind) have yielded light curves that are in reasonable agreement with observations \citep{bergerEA00, yostEA03, curranEA11, priceEA02}.  

Semi-analytic models developed by \citet{bergerEA03a}, \citet{frailEA05}, and \citet{vanderHorstEA05} have been used to estimate parameters of the GRB 030329 radio afterglow, including the isotropic equivalent energy $E_{k}$, the electron spectral index $p$, the electron and magnetic field energy fractions ($\epsilon_e$ and $\epsilon_B$, respectively), and the jet half-opening angle $\theta_j$, using broadband radio observations over the course of the first year.  \citet{vanderHorstEA08} extend this to the first $\sim3$ years.  We use all previously published observations of GRB 030329 between $0.84$ and $250$ GHz as well as archival VLA observations at 1.4, 4.9, and 8.5 GHz taken between 1.7 and 8.3 years after the burst to perform burst calorimetry deep into the non-relativistic regime.  The broadband afterglow is modelled using a method developed in \citet{meslerEA12a}.  

\section{GRB 030329}
At a redshift of $z = 0.1685$ \citep{greinerEA03}, the gamma-ray burst GRB 030329 is one of the closest GRBs detected to date.  Assuming a $\Lambda_\text{CDM}$ cosmology with $H_0 = 71$ km s$^{-1}$ Mpc$^{-1}$, $\Omega_M = 0.27$, and $\Omega_{\Lambda} = 0.73$, GRB 030329 is located at an angular distance of $d_A = 587$ Mpc, with 1.0 mas corresponding to 2.85 pc. The burst was first detected on March 29th, 2003 at 11:37 UTC by the High Energy Transient Explore 2 (\emph{HETE-2}) satellite, and was subsequently localized in the optical by \citet{petersonPrice03}.  The GRB 030329 radio afterglow was (and still is) the most luminous afterglow to ever have been observed, achieving a maximum flux density of $55$ mJy at $43$ GHz one week after the burst.  The relative proximity of the burst to the Earth, coupled with its extremely high luminosity, made it possible for the radio afterglow to be directly resolved by the Very Long Baseline Array (VLBA) and the Global VLBI Array at $5$ GHz \citep{taylorEA04, taylorEA05, pihlstromEA07, meslerEA12b}.  Additionally, the $5$ GHz radio afterglow was detected by the VLA for 8.3 years.  GRB 030329 has therefore provided a unique opportunity to study the evolution of both the luminosity and the physical size of a GRB afterglow


\section{Archival Data}

We compare our model to observations of the GRB 030329 radio afterglow luminosity. In addition to previously published data, we include late-period $1.4$, $4.9$, and $8.5$ GHz observations, taken from the NRAO data archive, that were made between 621 and 3018 days after the burst (see Table \ref{VLAobservations}).  The VLA data were reduced in the standard manner with the use of the AIPS software package.  Either 3C138 or 3C286 was used for absolute flux calibration, and J1051+213 or J1021+219 were used to determine the phase corrections.  

\begin{table*}[t]
\begin{center}
\begin{tabular}{|ccccc|}

\tableline 
    Date    & $\Delta$t & Frequency & Flux Density   & VLA Project Code \\
            &  (days)   &   (GHz)   &   ($\mu$Jy)    &                  \\
\tableline 
2004 Dec 09 &   621.2   &    1.4    & $650 \pm 70$   & AF414 \\
2004 Dec 09 	&   621.2   &    8.5    & $250 \pm 30$   & AF414 \\
2004 Dec 23 	&   635.1   &    1.4    & $590 \pm 70$   & AF414 \\
2004 Dec 23	&   635.1   &    4.9    & $370 \pm 60$   & AF414 \\
2004 Dec 23 	&   635.1   &    8.5    & $300 \pm 40$   & AF414 \\
2005 Jan 23	&   665.9   &    1.4    & $420 \pm 80$   & AK583 \\
2005 Jan 30 	&   672.9   &    1.4    & $480 \pm 70$   & AK583 \\
2005 Jan 31	&   673.7   &    1.4    & $510 \pm 80$   & AS796 \\
2005 Mar 31 	&   732.7   &    8.5    & $260 \pm 40$   & AK583 \\
2005 Apr 07	&   739.7   &    1.4    & $650 \pm 40$   & AF414 \\
2005 Apr 07	&   739.7   &    8.5    & $300 \pm 30$   & AF414 \\
2005 Jun 08 	&   801.5   &    1.4    & $380 \pm 10$   & AK583 \\
2005 Oct 21	&   937.0   &    8.5    & $100 \pm 30$   & AK583 \\
2005 Oct 29	&   944.5   &    8.5    & $120 \pm 30$   & AK583 \\
2005 Dec 14	&   990.5   &    8.5    & $180 \pm 20$   & AK583 \\
2005 Dec 26	&  1003.1   &    8.5    & $140 \pm 30$   & AK583 \\
2006 Mar 22	&  1088.7   &    1.4    & $630 \pm 30$   & AS864 \\
2006 Mar 22	&  1088.7   &    8.5    & $170 \pm 20$   & AS864 \\
2006 Apr 15	&  1266.2   &    8.5    & $80  \pm 30$   & AS864 \\
2006 Apr 29	&  1126.5   &    4.9    & $080 \pm 20$   & AS933 \\
2006 Sep 15	&  1266.2   &    4.9    & $170 \pm 50$   & AS864 \\
2011 Jul 03 &  3018.2   &    4.9    & $32  \pm 10$   & 10C-203 \\
\tableline 

\end{tabular}
\end{center}
\caption{Observations of the GRB 030329 radio afterglow at $1.4$, $4.9$, and $8.5$ GHz taken from the NRAO data archive and not appearing in a previous publication. \label{VLAobservations}}
\end{table*}   

To ensure the best possible fit of our semi-analytic model to the broadband afterglow, we also utilize previously-published measurements of the afterglow flux at $840$ MHz, $1.4$ GHz, $2.3$ GHz, $4.9$ GHz, $8.6$ GHz, $15$ GHz, $23$ GHz, $43$ GHz, $100$ GHz, and $250$ GHz.  Previously-published Westerbork Synthesis Radio Telescope (WSRT) data were taken from \citet{vanderHorstEA05} and \citet{vanderHorstEA08}.  Previously-published Very Large Array (VLA) data were taken from \citet{bergerEA03a}, \citet{frailEA05}, \citet{pihlstromEA07}, and \citet{meslerEA12b}.  The $100$ GHz and $250$ GHz data points are from \citet{shethEA03}.  The VLBI data points are from \citet{taylorEA04}, \citet{pihlstromEA07}, and \citet{meslerEA12b}


\section{Modeling the Jet Expansion}

In this section, we will discuss the semi-analytic method that we have developed for modelling GRB afterglow emission.  This model builds upon work appearing in \citet{meslerEA12a}. It will be outlined below with emphasis on improvements that have been made since its publication in \citet{meslerEA12a}.

Gamma ray bursts are modelled as initially relativistic, double-sided jets that propagate outward into an ambient circumburst medium.  In the following discussion, we will use primed quantities to refer to the reference frame that is comoving with the jet, we will use un-primed quantities with no subscript to refer to the reference frame in which the ISM is at rest, and we will use quantities with the subscript $\earth$ to refer to the reference frame of an Earthbound observer.  

\subsection{Jet Hydrodynamics}

An expression for the evolution of the jet Lorentz factor $\Gamma$ can be found by invoking the requirement for the conservation of energy.  The jet will sweep up material from the circumburst medium as it propagates, forcing the jet to decelerate.  \citet{peer12} shows that

\begin{equation}
\frac{d\Gamma}{dm} = -\frac{\hat{\gamma}\left(\Gamma^2 - 1\right) - \left(\hat{\gamma} - 1\right)\Gamma\beta^2}{M_\text{ej} + \epsilon m + (1-\epsilon)m\left[2\hat{\gamma}\Gamma - \left(\hat{\gamma} - 1\right)\left(1 + \Gamma^{-2}\right)\right]},
\label{GammaEquation}
\end{equation}

\noindent where $\Gamma$ is the Lorentz factor of the jet, $M_\text{ej}$ is the initial mass of the jet ejecta, $m$ is the total mass that has been swept up by the jet, $\beta = \left(1-\Gamma^{-2}\right)^{1/2}$ is the normalized bulk velocity, $\hat{\gamma} \simeq (4\Gamma+1)/(3\Gamma)$ is the adiabatic index \citep{huangEA00}, and 

\begin{equation}
\epsilon = \epsilon_e\frac{{t'}_\text{syn}^{-1}}{{t'}_\text{syn}^{-1} + {t'}_\text{ex})^{-1}}
\end{equation}

\noindent is the radiative efficiency of the jet \citep{daiEA99}.  The expansion time $t'_\text{ex} = t/(\beta\Gamma c)$ and the synchrotron cooling time $t'_\text{syn} = 6\pi c/\sigma_T\epsilon_e\Gamma m_pB'^2$ \citep{daiLu98b}, where $c$ is the speed of light, $\sigma_T$ is the Thompson scattering cross-section, $\epsilon_e$ is the fraction of the burst energy stored in the electrons, $m_p$ is the proton mass, and $B'$ is the magnitude of the comoving frame magnetic field.  

Our semi-analytic method is capable of producing light curves in arbitrary density profiles, but we limit ourselves for the sake of simplicity to a uniform density typical of an interstellar medium (ISM) and an $n(r) \propto r^{-2}$ stellar wind.  Density profiles are modelled as a series of uniform grid points of width $10^{-4}$ pc.  The model is fed an initial kinetic energy $E_K$, ejecta mass $M_\text{ej}$, and jet half-opening angle $\theta_0$.  The total mass swept up by the jet ($m$), the jet's Lorentz factor ($\Gamma$), the width of the leading edge of the jet perpendicular to the observer's line of sight ($a$), and the isotropic frame time ($t$) are then solved for simultaneously as a function of the radius ($r$) of the jet. 

The evolution of the jet is identical to the case where the GRB outflow is isotropic until $t = t_\text{jet}$, when the center of the jet comes into causal contact with its edge.  For a jet with half-opening angle $\theta_j = \arctan{a/r}$, this occurs when $\Gamma \simeq 1/\theta_j$.  At  $t_\text{jet}$, the jet experiences rapid lateral expansion, leading to an increase in the amount of circumburst material being swept up by the jet with time.  The increase in swept-mass leads to a faster deceleration and a decrease in the jet luminosity.  Eventually, the jet is decelerated to the point that it is no longer relativistic, and it transitions to spherical expansion.  The transition time to non-relativistic expansion is denoted $t_\text{NR}$.

\subsection{Determining the Initial Lorentz Factor} 

GRB 030329 is unique in that its afterglow has been directly resolved using VLBI at 5 GHz \citep{taylorEA04,taylorEA05,pihlstromEA07,meslerEA12b}. With multiple epochs of observation, the expansion history of the burst can be determined directly, allowing us to place constraints on the initial Lorentz factor of the burst.  

Initially, the mass being swept up by the GRB jet will be negligible as compared to the initial jet ejecta mass $M_\text{ej}$.  The jet will coast at nearly constant speed in this regime assuming that $\epsilon \ll 1 \ll M_\text{ej}/m$.  It is only when the sum of the two terms in the denominator of equation \ref{GammaEquation} that are dependent upon the swept mass $m$ is of the same order as the initial ejecta mass $M_\text{ej}$ that the jet begins to decelerate appreciably.  We will define the isotropic frame time $t = t_\text{dec}$ as the time at which

\begin{equation}
M_\text{ej} = \epsilon m + \left(1-\epsilon\right)m\left[2\hat{\gamma}\Gamma - \left(\hat{\gamma}-1\right)\left(1+\Gamma^{-2}\right)\right].
\label{tDecEquation}
\end{equation}

The afterglow linear size evolves according to the comoving sound speed \citep{huangEA00}

\begin{equation}
c'_s = \sqrt{\frac{\hat{\gamma}\left(\hat{\gamma}-1\right)\left(\Gamma-1\right)}{1+\hat{\gamma}\left(\Gamma-1\right)}}c.
\end{equation}

\noindent In the regime where $t < t_\text{dec}$, the Lorentz factor is nearly constant, meaning that the afterglow expands laterally at a nearly constant rate $\beta_\perp$.  Using the relationship between the isotropic frame time and the Earth frame time

\begin{equation}
dt = \Gamma dt' = \Gamma\left(\Gamma + \sqrt{\Gamma^2 - 1}\right)\ dt_\Earth \label{earthFrameTimeEqn},
\end{equation}  

\noindent we find that the relationship between the expansion rate $\beta_\perp$ of the afterglow and the initial Lorentz factor is

\begin{equation}
\beta_\perp \simeq \frac{\sqrt{3}}{3}\left(\Gamma_0+\sqrt{\Gamma_0^2-1}\right)\sqrt{\frac{4\Gamma_0^3+\Gamma_0^2-4\Gamma_0-1}{\Gamma_0\left(4\Gamma_0^2-1\right)}}
\label{betaPerpEquation}
\end{equation}

\noindent for $t < t_\text{dec}$. 

\begin{figure}[ht]
\centering
\includegraphics[width=0.45\textwidth]{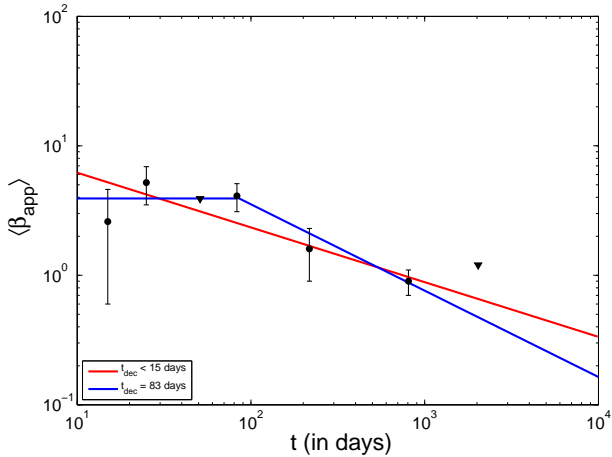}
\caption{The observed transverse expansion rate $\langle\beta_\text{app}\rangle$.  The red line corresponds to a single power law with index $-0.42$ as in \citet{meslerEA12b}.  The blue line is a piecewise function consisting of a constant value before a time $t_\text{dec} = 83$ days and a single power law thereafter.  \label{expansionRateFig}}
\end{figure}

The average apparent expansion rate of the afterglow (in units of the speed of light) is defined as 

\begin{equation}
\langle \beta_{\rm app} \rangle = \frac{(1+z)R_\perp}{ct_\Earth},
\end{equation}

\noindent where $R_\perp$ is the physical radius of the image, $z$ is the source's cosmological redshift, $t$ is the Earth-frame time of observation, and $c$ is the speed of light.  The observed average lateral expansion rate of the GRB 030329 radio afterglow is shown in Fig. \ref{expansionRateFig}.  At early times ($t < t_\text{dec}$), the lateral expansion rate is nearly constant, and, therefore, $\beta_\perp \simeq \langle\beta_\text{app}\rangle$.  After $t = t_\text{dec}$, however, the jet begins to decelerate and both $\beta_\perp$ and $\langle\beta_\text{app}\rangle$ begin to decrease.  The time $t_\text{dec}$ will therefore show up in the plot of $\langle\beta_\text{app}\rangle$ vs. $t$ as a break where $\frac{d\langle\beta_\text{app}\rangle}{dt}$ begins to decrease from its initial value of $\sim0$. 

Figure \ref{expansionRateFig} shows two fits to the expansion history of the burst.  The first fit is a simple power law with index $-0.42$, as in \citet{meslerEA12b}.  In this interpretation, the time $t_\text{dec}$ occurs before the time of the first VLBI observation at day 15, and we do not see the coasting phase during which the jet is moving at a nearly constant velocity.  In the second model, the jet coasts at a constant rate until day $t_\text{dec} = 83$ days and then begins to decelerate.  Because both models fit the data equally well, it is not possible to distinguish between them.  Models of the average apparent expansion rate which assume $t_\text{dec} > 83$ days, however, produce steadily worse fits as $t_\text{dec}$ is increased, so we argue that $t_\text{dec} \lesssim 83$ days.  The lower bound on the possible values of $t_\text{dec}$ can be found by turning to our MHD models.  An early $t_\text{dec}$ implies a high initial Lorentz factor and a longer transition time between the coasting phase and the decelerating phase.  In order for $t_\text{dec}$ to have occurred before the date of the first observation at day 15, the jet must have had a low enough initial Lorentz factor for its average apparent expansion rate to be adequately modelled as a single power law from day 15 onward.  Our MHD models produce jets with average apparent expansion rates that can be modelled as single power laws after day 15 only if $t_\text{dec} \gtrsim 1$ day.  For a range in $t_\text{dec}$ of $1 \lesssim t_\text{dec} \lesssim 83$ days, we obtain $4.5 \lesssim \langle\beta_\text{app}\rangle \lesssim 7.0$.  Using Eqn \ref{betaPerpEquation}, we then obtain an estimate for the initial Lorentz factor of $4 \lesssim \Gamma_0 \lesssim 6$. 

The afterglow emission from GRB 030329 has previously been interpreted as coming from a two-component jet \citep{bergerEA03a, frailEA05}.  One of the components was highly relativistic with a small initial half-opening angle of $\theta_0 \simeq 5^\circ$ and was responsible for the high energy emission (optical and higher frequencies), while the other component was only mildly relativistic with a larger jet half-opening angle $\theta_0 \simeq 17^\circ$, and was responsible for emission at frequencies in the optical and below.  Our value of $4 \lesssim \Gamma_0 \lesssim 6$ is consistent with the Lorentz factor expected for the wider, moderately relativistic jet component.   

The initial Lorentz factor is related to the ejecta mass $M_\text{ej}$ and the kinetic energy $E_k$ via

\begin{equation}
E_k = \left(\Gamma_0 - 1\right)M_\text{ej}c^2.
\label{kineticEnergyEquation}
\end{equation}

\noindent Determination of the initial Lorentz factor can therefore provide an important constraint on the ratio between the initial burst kinetic energy and the ejecta mass.  

\subsection{Fitting the $\chi_r^2$ MHD Models}

\begin{figure*}[ht]
\centering
\includegraphics[width=0.6\textwidth]{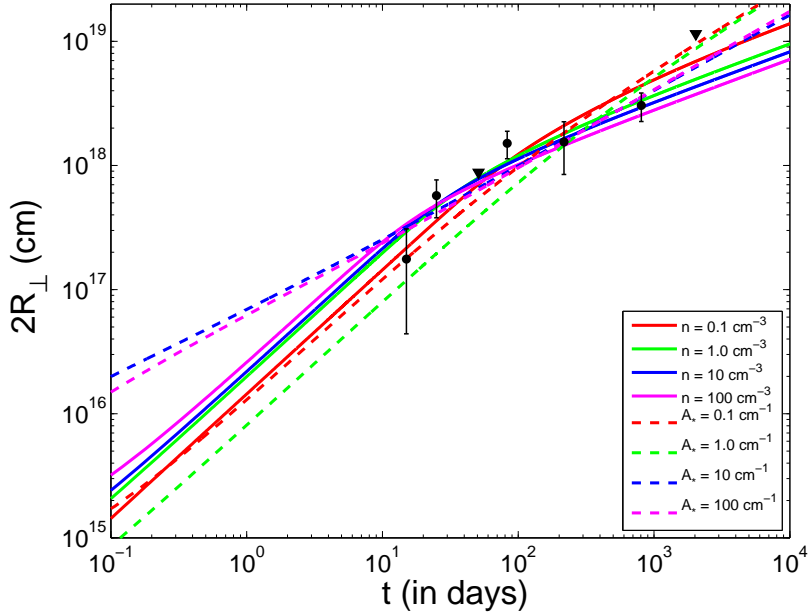}
\caption{Fits of the GRB 030329 radio afterglow linear size to MHD models.  Solid lines refer to models in which a uniform medium was assumed, while dashed lines indicate a stellar wind.   \label{MHDfitFig}}
\end{figure*}

Using the observations of the burst linear size (Fig. \ref{expansionRateFig}), we produce a best-fit to our MHD models by varying the kinetic energy $E_k$, the ejecta mass $M_\text{ej}$, and the initial jet half-opening angle $\theta_0$.  Best-fits to a variety of wind and uniform density profiles are shown in Fig. \ref{MHDfitFig}.  We find that the afterglow size evolution can be successfully fitted to either a wind or a uniform density, but that lower values of $\chi_r^2 = \chi^2/N$, where $N$ is the number of degrees of freedom present in the model, can be obtained for fits of uniform media to the afterglow size evolution than for stellar winds ($\chi_{r\text{, uniform}}^2 \gtrsim 1.2$ versus $\chi_{r\text{, wind}}^2 \gtrsim 2.2$).  The medium surrounding GRB 030329 is therefore perhaps more characteristic of a uniform density ISM than a stellar wind.  A good fit can be obtained for $2 \lesssim \Gamma_0 \lesssim 10$ over the entire range in $n$ and $A_*$ that we searched ($10^{-1} < n < 10^2$ cm$^{-3}$ and $0.3 < A_* < 6$ cm$^{-1}$).   

\section{Modelling the Light Curve}

While the linear size evolution of the afterglow can give us insight into the initial Lorentz factor of the jet, it can only be used to determine the ratio between the initial kinetic energy and the ejecta mass if the nature of the circumburst medium is unknown.  In order to determine the values of the kinetic energy and the ejecta mass, as well as the values of the other burst parameters, we must expand our model to include the properties of the afterglow emission.  We build upon the model detailed in section 5 of \citet{meslerEA12a} to model the GRB 030329 synchrotron emission by fitting to broadband radio afterglow observations between $840$ MHz and $250$ GHz taken between 1 and 3018 days after the burst. 

\subsection{Spherical Emission and Beaming}

The leading edge of a GRB jet is not a flat plane that propagates directly toward an Earthbound observer.  Rather, it is a section of a spherical surface, meaning that material at different positions along the leading edge of the jet has different velocity components toward the observer.   Relativistic beaming will cause radiation that is emitted by material with the largest velocity component in the direction of the observer to be beamed toward the observer more than radiation that is emitted by material with a slightly smaller velocity component.  Additionally, jet material at different locations in the jet is not all at a uniform distance from the observer, leading to differences in light travel time throughout the jet.  The overall effect of the relativistic beaming on the afterglow is an increase in the total observed flux at early times when the jet is still relativistic.  The difference in arrival times of photons emitted from different regions of the jet makes the light curve broader and more smooth than it would be otherwise.

To incorporate the effects of spherical emission and beaming into our model, equations 27-32 of \citet{meslerEA12a} must be modified to account for the fact that radiation emitted by different parts of the jet at the same time $t_\text{emit}$ will not all reach the observer at the same time $t_\earth$, nor will it all be beamed at the observer to the same degree.

The arrival time $t_\earth$ at which radiation emitted by the jet reaches the observer will be dependent upon the angle $\theta$ at which it was emitted with respect to a line connecting the center of the GRB progenitor to the observer.  Material at angle $\theta = 0$ will arrive at the observer first, while material emitted at $\theta = \theta_j$ will arrive last.  The equation for the equal arrival time surface is

\begin{equation}
t_\earth = \int{\frac{\left(1-\beta\cos{\theta}\right)}{c\beta}\ dr} = \text{const}.
\label{equalArrivalTimeEquation}
\end{equation}

Numerical integration of equation \ref{equalArrivalTimeEquation} yields the radius at which emission at some angle $\theta$ was emitted in order to reach the observer at time $t_\earth$.  To account for the time and angular dependence of the beaming of radiation toward the observer, we integrate up the luminosities of the material at each location [r, $\theta$, $\phi$] which emit radiation that arrives at the observer at time $t_\earth$ to find the total afterglow flux density:

\begin{equation}
F_{\nu,\earth} = \frac{1}{4\pi D^2} \iint\limits_{\Omega_j}{\frac{L'_{\nu'}[r(\theta)]\mathcal{D}^3}{\Omega_j}d\cos{\theta}\ d\phi},
\label{correctedFluxEquation}
\end{equation}

where $\mathcal{D}$ is the Doppler factor $\mathcal{D} = 1/{\Gamma(1-\beta\cos{\theta})}$ and $\Omega_j = 2\pi(1 - \cos{\theta_j})$ is the solid angle occupied by the jet \citep{moderskiEA00}.  The quantity $L'_{\nu'}$ is the comoving frame luminosity of the afterglow, which can be determined via equations 27-32 of \citet{meslerEA12a}. 


\section{Burst Calorimetry From the Broadband Afterglow}

The MHD and emission models detailed above were used to perform calorimetry on the broadband radio afterglow of GRB 030329.  Seven parameters were fit simultaneously: the kinetic energy ($E_k$), the ejecta mass ($M_\text{ej})$, the jet half-opening angle ($\theta_j$), the fraction of the total burst energy stored in magnetic fields ($\epsilon_B$) and in electrons ($\epsilon_e$), the electron power law index ($p$), and the medium density (n) in the case of a uniform medium or the wind density scaling factor ($A_*$) in the case of a wind.  The factor $A_*$ is set such that the medium density $\rho(r) = 5\times10^{11}A_*r^{-2}$ g cm$^{-1}$ for a stellar wind.    

\begin{figure*}[ht]
\centering
\includegraphics[width=\textwidth]{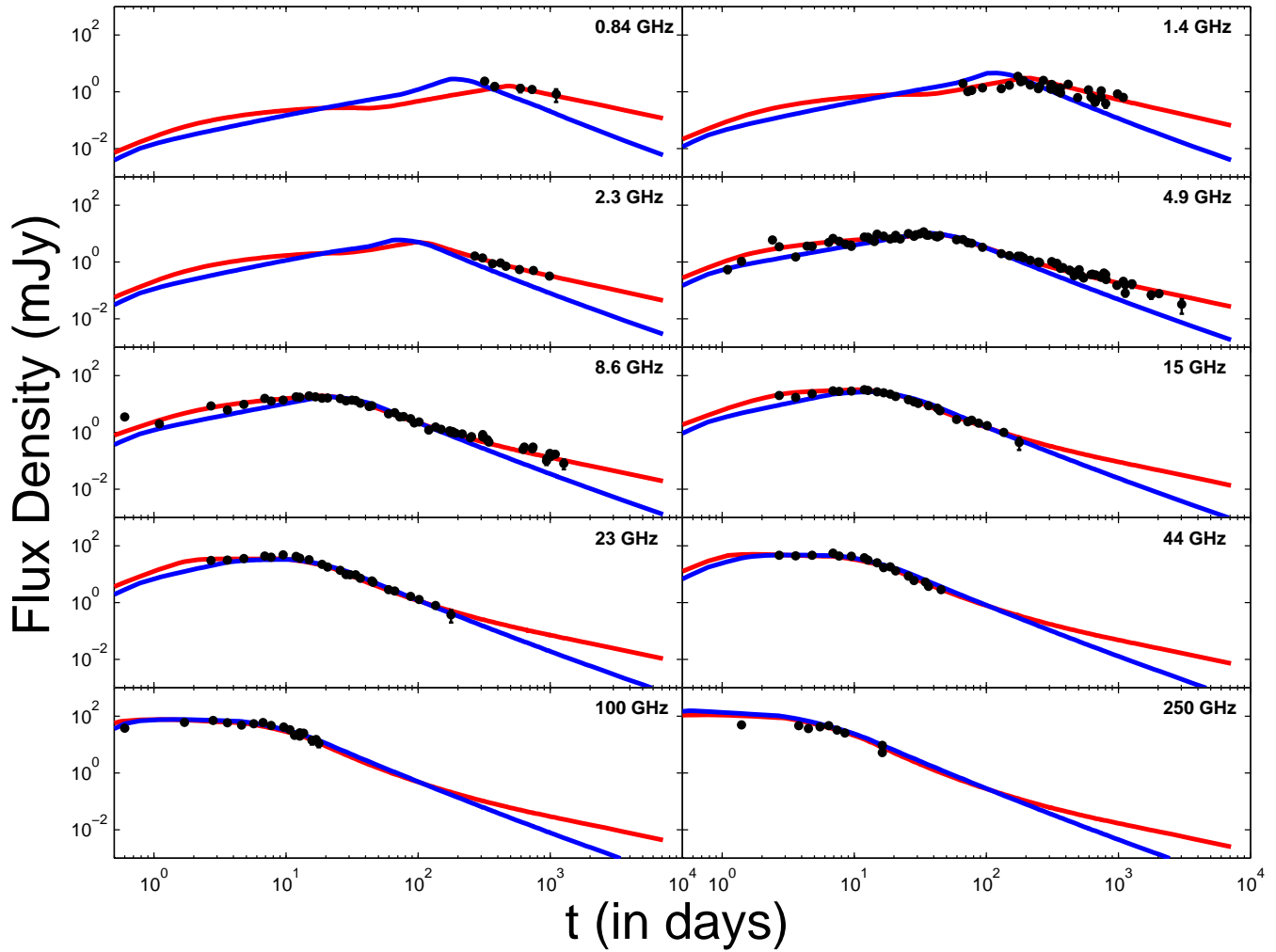}
\caption{Best fits to the GRB 030329 broadband radio afterglow assuming a uniform medium (red line) and a stellar wind (blue line).  \label{lightCurveFitFig}}
\end{figure*}

The values that are obtained for the GRB 030329 burst parameters appear in Table \ref{lightCurveFitTable}.  Synthetic light curves produced using the best fit uniform medium and wind models are shown in Fig. \ref{lightCurveFitFig}.  We incorporate the \citet{goodman97} model of interstellar scintillation to account for the time-dependent scatter of the low-frequency ($\lesssim 10$ GHz) data.  The high-frequency data ($\gtrsim 10$ GHz) exhibits mild departures from the expected smooth behaviour of the afterglow emission, possibly due to slight clumping of the circumburst medium.   The large values of $\chi_r^2$ are due to these mild departures from smooth behaviour.  Fitting only the data at or below $4.9$ GHz yields $\chi_r^2 = 1.9$ for the best-fit uniform medium and $\chi_r^2 = 24$ for the best-fit wind.

\begin{table*}[t]
\begin{center}
\begin{tabular}{|ccccccccc|}

\tableline 
Medium Type   &     $E_k$       &    $M_\text{ej}$    & $\theta_0$ & $\epsilon_B$ & $\epsilon_e$ &  $p$  & $A^1$ & $\chi_r^2$ \\
              & ($10^{51}$ erg) &     ($M_\sun$)      & ($^\circ$) &              &              &       &       &            \\
\tableline 
uniform (ISM) &     $1.4$       & $2.2\times 10^{-4}$ &    $24$    &   $0.046$    &     $0.33$   & $2.2$ & $6.5$ &   $16.8$    \\
wind          &     $1.5$       & $9.0\times 10^{-5}$ &    $37$    &   $0.15$     &     $0.33$   & $2.2$ & $0.8$ &   $36.0$    \\
\tableline 

\end{tabular}
\end{center}
\caption{Best fits to the GRB 030329 burst parameters assuming either a uniform density or a stellar wind circumburst medium.  The medium density scaling parameter$^1$ $A = n$ for a uniform medium and $A = A_*$ for a wind.  Note that the values of $E_k$ listed here correspond to the kinetic energy of the jet at the time of the burst.  \label{lightCurveFitTable}}
\end{table*}   

The best-fit uniform density and wind models are both capable of reproducing the broadband afterglow for the first $\sim300$ days.  The transition to non-relativistic expansion begins at approximately 42 days.  The flux will decay more quickly in the case of a stellar wind than in the case of a uniform medium after the non-relativistic transition time $t_\text{NR}$.  The scatter of the data due to refractive scintillation and large-scale diffractive effects means that the shallower decay due to a uniform medium is not obvious until several hundred days after the burst (Fig. \ref{lightCurveFitFig}).   It is clear that a uniform medium is preferred over a stellar wind, however, by looking at the data that was obtained after $\sim600$ days.  This finding is consistent with previous work \citep{bergerEA03a, frailEA05, vanderHorstEA05, vanderHorstEA08, meslerEA12b}.  A stellar wind model can be made to better fit the broadband afterglow emission if the wind scaling parameter is increased to $A_* > 1$.  In doing so, however, the initial Lorentz factor must also be increased in order to keep the jet from becoming non-relativistic too early.  A larger value of $A_*$, then, will provide a better fit to the broadband afterglow at the expense of a poorer fit to the afterglow expansion rate if $E_\text{k}$, $M_\text{ej}$, and $\theta_j$ are held constant.

Our fitted burst parameters are broadly consistent with previous work \citep{bergerEA03a, frailEA05, vanderHorstEA05, vanderHorstEA08} (Table \ref{modelComparisonTable}).  We find that the initial burst kinetic energy is $E_k = 1.4\times10^{51}$ erg.  Radiative losses and energy losses do to adiabatic expansion reduce this to $E_k = 0.59\times10^{51}$ ergs by $t_\text{jet} = 13$ days.  The ejecta mass is found to be $M_\text{ej} = 2.2\times10^{-4} M_\sun$, yielding $\Gamma_0 = 4.5$.  We also find that $\theta_0 = 24^\circ$, $\epsilon_B = 0.045$, $\epsilon_e = 0.33$, and $p = 2.2$.  The value we obtain for the medium density of $n = 6.5$ cm$^{-3}$ is somewhat larger than has been found by previous authors.  The wide range in densities obtained for the GRB 030329 circumburst medium is probably due to differences in the individual models employed by each author.  Uncertainties in the structure of the jet magnetic fields and the time dependence of $\epsilon_B$ and $\epsilon_e$ limit the accuracy of any gamma ray burst emission model.  Given the imperfect nature of the current understanding of the jet physics, we estimate that the values we obtain for the burst parameters are accurate to within approximately a factor of $1.5$.

\begin{table*}[t]
\begin{center}
\begin{tabular}{|ccccccccc|}

\tableline 
Model                &  $t_j$  & $t_\text{NR}$ &     $E_k$       & $\theta_0$ & $\epsilon_B$ & $\epsilon_e$ &  $p$  & $n$    \\
                     & (days)  &     (days)    & ($10^{51}$ erg) & ($^\circ$) &              &              &       &        \\
\tableline 
Relativistic$^1$     & $10$    &      N/A      &     $0.67$      &    $26$    &   $0.042$    &     $0.19$   & $2.2$ & $3.0$  \\
Full$^2$             & $14$    &      $48$     &     $0.90$      &    $26$    &   $0.074$    &     $0.17$   & $2.2$ & $2.2$  \\
Non-Relativistic$^2$ & N/A     &      $50$     &     $0.78$      &    N/A     &   $0.13$     &     $0.06$   & $2.2$ & $1.3$  \\
Relativistic$^3$     & $10$    &      N/A      &     $0.24$      &    $42$    &   $0.43$     &     $0.28$   & $2.2$ & $0.8$  \\
Non-Relativistic$^4$ & N/A     &      $80$     &     $0.34$      &    N/A     &   $0.49$     &     $0.25$   & $2.2$ & $0.8$  \\
Full                 & $13$    &      $42$     &     $0.59$      &    $24$    &   $0.045$    &     $0.33$   & $2.2$ & $6.5$  \\
\tableline 

\end{tabular}
\end{center}
\caption{Comparison of best-fit parameters for the various models that have been produced for the GRB 030329 radio afterglow.  These models were first published in \citet{bergerEA03a}$^1$, \citet{frailEA05}$^2$, \citet{vanderHorstEA05}$^3$, and \citet{vanderHorstEA08}$^4$.  The bottommost model is the best-fit uniform density model presented in this work.  Note that the kinetic energies listed here are valid at $t = t_\text{jet}$ for models valid in the relativistic regime and at $t = t_\text{NR}$ for models that are valid solely in the non-relativistic regime. \label{modelComparisonTable}}
\end{table*}  

\begin{figure}[ht]
\centering
\includegraphics[width=0.45\textwidth]{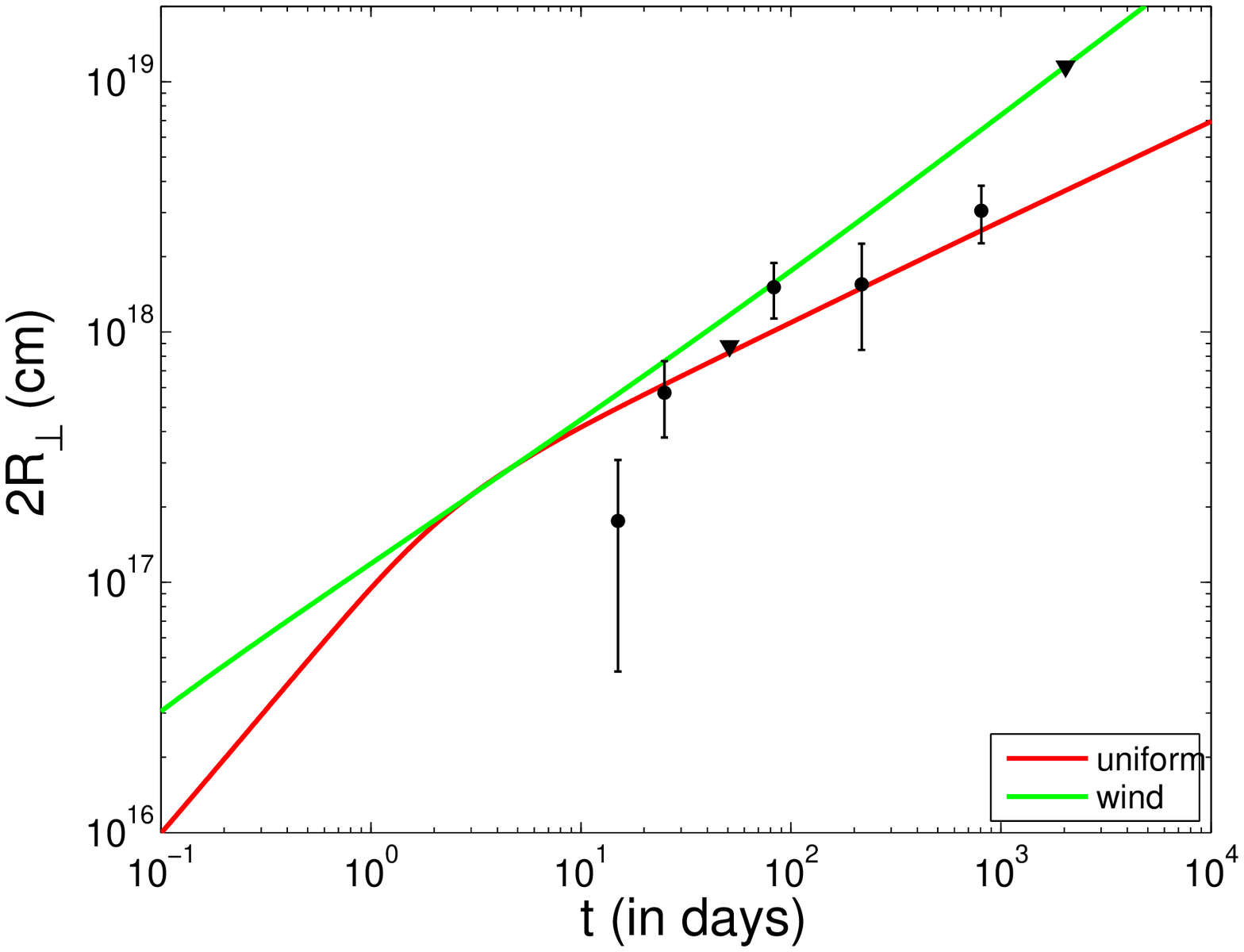}
\caption{Fits of the uniform and wind density models from Table \ref{lightCurveFitTable} to the observed apparent lateral afterglow expansion rate $\langle\beta_\text{app}\rangle$.   \label{LCmodelMHDfitFig}}
\end{figure}

The uniform medium and wind models that produced the best fit to the broadband afterglow were themselves fit to the observed radio afterglow image size evolution (Fig. \ref{LCmodelMHDfitFig}).  Both models fit the evolution of the afterglow size, though the uniform medium fits significantly better ($\chi_\text{r, uniform}^2 = 1.1$ vs $\chi_\text{r, wind}^2 = 3.9$).  From the initial kinetic energy and the ejecta mass, we find that $\Gamma_0 = 4.5$ in the uniform medium case and $\Gamma_0 = 10.3$ in the stellar wind case.  The uniform medium model initial Lorentz factor agrees with our estimate of the initial Lorentz factor from section 2.2 of $4 \lesssim \Gamma_0 \lesssim 6$, while the wind model initial Lorentz factor does not.    

\section{Conclusions}

We have presented the first simultaneous fit of a gamma-ray burst afterglow model to a GRB's broadband light curve and its observed afterglow expansion rate.  More than eight years of afterglow observations at radio frequencies between 840 MHz and 250 GHz were employed to perform accurate calorimetry on the gamma-ray burst GRB 030329 deep into the non-relativistic phase.  Values for the burst parameters were determined, and the nature of the circumburst medium was explored.  

By noting that a GRB jet will coast at a nearly constant velocity until a time $t_\text{dec}$ where it has begun to sweep up a significant amount of circumburst material, we derive a relationship between the average apparent afterglow lateral expansion rate, $\langle\beta_\text{app}\rangle$, and the initial Lorentz factor $\Gamma_0$.  The initial Lorentz factor for GRB 030329 is found to be $4 \lesssim \Gamma_0 \lesssim 6$.  \citet{bergerEA03a} found that the GRB 030329 radio afterglow was best modelled as the product of a jet component that was initially mildly-relativistic.  This interpretation has been subsequently supported by other authors \citep{frailEA05,vanderHorstEA05,vanderHorstEA08}, and is in agreement with our estimate of $\Gamma_0$.  

The flux falls off more slowly in the case of a uniform medium than in the case of a wind once the jet becomes non-relativistic, so late-time observations of a GRB afterglow provide important insight into the nature of the circumburst medium.  Refractive scintillation and large-scale diffractive effects produce a significant scatter in the data at $\lesssim 10$ GHz, meaning that observations of the flux at very late times, when the predicted afterglow evolution is significantly different for a wind than for a uniform medium, are extremely valuable in determining the nature of the circumburst medium.  Our light curve models incorporate data out to 3018 days after the burst, allowing us to better determine the burst parameters and the nature of the circumburst medium than has been previously possible for GRB 030329.  

We find that the best fit of a uniform density model to the broadband radio afterglow predicts values for the burst parameters that are similar to the values that are obtained when a stellar wind density profile is assumed.  The goodness of fit $\chi_r^2$ of the model that assumes a uniform density medium, however, is better by a factor of $\sim2.1$ than the goodness of fit of the wind model, suggesting that a uniform medium is preferred over a stellar wind.  This is in agreement with previous work \citep{bergerEA03a, frailEA05, meslerEA12b}, but runs contrary to what is expected if GRBs have stellar progenitors.  

The best-fit wind and uniform density models were fitted to the observed average apparent afterglow lateral expansion rate and are both in good agreement with the data.  The uniform density model agrees nicely with our estimate of $4 \lesssim \Gamma_0 \lesssim 6$, while the wind model does not, again suggesting that the circumburst medium is more characteristic of a uniform ISM than a stellar wind.

The ability to determine the initial Lorentz factor of a GRB jet provides a powerful constraint on models of the burst evolution.  The initial Lorentz factor of the burst determines the time $t_\text{dec}$ when the jet transitions to non-relativistic expansion as well as the jet's initial lateral expansion rate.  By fitting the afterglow expansion history to determine $\Gamma_0$, we limit the parameter space from which we can build a model of the broadband afterglow.  In the case of GRB 030329, this means that no stellar wind environment remains that can be used as a suitable model of the circumburst medium.  In order to determine the initial Lorentz factor, however, we must have enough observations of the afterglow image size to determine $t_\text{dec}$.  The GRB 030329 afterglow is currently the only gamma ray burst afterglow to have been directly resolved.  In the future, more luminous, low-redshift bursts will need to be imaged with VLBI so that combined afterglow and lateral expansion evolution fitting can be applied beyond the case of GRB 030329.


\begin{acknowledgements}The National Radio Astronomy Observatory is a facility of the National Science Foundation operated under cooperative agreement by Associated Universities, Inc.
\end{acknowledgements}




\begin{thebibliography}{32}
\expandafter\ifx\csname natexlab\endcsname\relax\def\natexlab#1{#1}\fi

\bibitem[{Berger {et~al.}(2003)Berger, Kulkarni, Pooley, Frail, McIntyre, Wark,
  Sari, Soderberg, Fox, Yost, \& Price}]{bergerEA03a}
Berger, E., Kulkarni, S.~R., Pooley, G., Frail, D.~A., McIntyre, V., Wark,
  R.~M., Sari, R., Soderberg, A.~M., Fox, D.~W., Yost, S., \& Price, P.~A.
  2003, \nat, 426, 154

\bibitem[{Berger {et~al.}(2000)Berger, Sari, Frail, Kulkarni, Bertoldi, Peck,
  Menten, Shepherd, Moriarty-Schieven, Pooley, Bloom, Diercks, Galama, \&
  Hurley}]{bergerEA00}
Berger, E., Sari, R., Frail, D.~A., Kulkarni, S.~R., Bertoldi, F., Peck, A.~B.,
  Menten, K.~M., Shepherd, D.~S., Moriarty-Schieven, G.~H., Pooley, G., Bloom,
  J.~S., Diercks, A., Galama, T.~J., \& Hurley, K. 2000, \apj, 545, 56

\bibitem[{Chevalier \& Li(2000)}]{chevalierLi00}
Chevalier, R.~A. \& Li, Z.-Y. 2000, \apj, 536, 195

\bibitem[{Curran {et~al.}(2011)Curran, Starling, van~der Horst, Wijers,
  de~Pasquale, \& Page}]{curranEA11}
Curran, P.~A., Starling, R.~L.~C., van~der Horst, A.~J., Wijers, R.~A.~M.~J.,
  de~Pasquale, M., \& Page, M. 2011, Advances in Space Research, 47, 1362

\bibitem[{{Dai} {et~al.}(1999){Dai}, {Huang}, \& {Lu}}]{daiEA99}
{Dai}, Z.~G., {Huang}, Y.~F., \& {Lu}, T. 1999, \apj, 520, 634

\bibitem[{Dai \& Lu(1998)}]{daiLu98b}
Dai, Z.~G. \& Lu, T. 1998, \mnras, 298, 87

\bibitem[{{Fenimore} {et~al.}(1996){Fenimore}, {Madras}, \&
  {Nayakshin}}]{fenimoreEA96}
{Fenimore}, E.~E., {Madras}, C.~D., \& {Nayakshin}, S. 1996, \apj, 473, 998

\bibitem[{Frail {et~al.}(2005)Frail, Soderberg, Kulkarni, Berger, Yost, Fox, \&
  Harrison}]{frailEA05}
Frail, D.~A., Soderberg, A.~M., Kulkarni, S.~R., Berger, E., Yost, S., Fox,
  D.~W., \& Harrison, F.~A. 2005, \apj, 619, 994

\bibitem[{{Goodman}(1997)}]{goodman97}
{Goodman}, J. 1997, New Astronomy, 2, 449

\bibitem[{Granot(2007)}]{granot07}
Granot, J. 2007, in Revista Mexicana de Astronomia y Astrofisica Conference
  Series, Vol.~27, Revista Mexicana de Astronomia y Astrofisica, vol. 27,
  140--165

\bibitem[{Granot \& Sari(2002)}]{granotSari02}
Granot, J. \& Sari, R. 2002, \apj, 568, 820

\bibitem[{Greiner {et~al.}(2003)Greiner, Peimbert, Estaban, Kaufer, Jaunsen,
  Smoke, Klose, \& Reimer}]{greinerEA03}
Greiner, J., Peimbert, M., Estaban, C., Kaufer, A., Jaunsen, A., Smoke, J.,
  Klose, S., \& Reimer, O. 2003, GRB Coordinates Network, 2020, 1

\bibitem[{{Huang} {et~al.}(2000){Huang}, {Gou}, {Dai}, \& {Lu}}]{huangEA00}
{Huang}, Y.~F., {Gou}, L.~J., {Dai}, Z.~G., \& {Lu}, T. 2000, \apj, 543, 90

\bibitem[{{Katz}(1994)}]{katz94}
{Katz}, J.~I. 1994, \apjl, 432, L107

\bibitem[{{Mesler} {et~al.}(2012{\natexlab{a}}){Mesler}, {Pihlstr{\"o}m},
  {Taylor}, \& {Granot}}]{meslerEA12b}
{Mesler}, R.~A., {Pihlstr{\"o}m}, Y.~M., {Taylor}, G.~B., \& {Granot}, J.
  2012{\natexlab{a}}, \apj, 759, 4

\bibitem[{{Mesler} {et~al.}(2012{\natexlab{b}}){Mesler}, {Whalen},
  {Lloyd-Ronning}, {Fryer}, \& {Pihlstr{\"o}m}}]{meslerEA12a}
{Mesler}, R.~A., {Whalen}, D.~J., {Lloyd-Ronning}, N.~M., {Fryer}, C.~L., \&
  {Pihlstr{\"o}m}, Y.~M. 2012{\natexlab{b}}, \apj, 757, 117

\bibitem[{M{\'e}sz{\'a}ros(2002)}]{meszaros02}
M{\'e}sz{\'a}ros, P. 2002, \araa, 40, 137

\bibitem[{{Meszaros} \& {Rees}(1993)}]{meszarosRees93}
{Meszaros}, P. \& {Rees}, M.~J. 1993, \apj, 405, 278

\bibitem[{{Moderski} {et~al.}(2000){Moderski}, {Sikora}, \&
  {Bulik}}]{moderskiEA00}
{Moderski}, R., {Sikora}, M., \& {Bulik}, T. 2000, \apj, 529, 151

\bibitem[{{Pe'er}(2012)}]{peer12}
{Pe'er}, A. 2012, \apjl, 752, L8

\bibitem[{Peterson \& Price(2003)}]{petersonPrice03}
Peterson, B.~A. \& Price, P.~A. 2003, GRB Coordinates Network, 1985, 1

\bibitem[{Pihlstr{\"o}m {et~al.}(2007)Pihlstr{\"o}m, Taylor, Granot, \&
  Doeleman}]{pihlstromEA07}
Pihlstr{\"o}m, Y.~M., Taylor, G.~B., Granot, J., \& Doeleman, S. 2007, \apj,
  664, 411

\bibitem[{Price {et~al.}(2002)Price, Berger, Reichart, Kulkarni, Yost,
  Subrahmanyan, Wark, Wieringa, Frail, Bailey, Boyle, Corbett, Gunn, Ryder,
  Seymour, Koviak, McCarthy, Phillips, Axelrod, Bloom, Djorgovski, Fox, Galama,
  Harrison, Hurley, Sari, Schmidt, Brown, Cline, Frontera, Guidorzi, \&
  Montanari}]{priceEA02}
Price, P.~A., Berger, E., Reichart, D.~E., Kulkarni, S.~R., Yost, S.~A.,
  Subrahmanyan, R., Wark, R.~M., Wieringa, M.~H., Frail, D.~A., Bailey, J.,
  Boyle, B., Corbett, E., Gunn, K., Ryder, S.~D., Seymour, N., Koviak, K.,
  McCarthy, P., Phillips, M., Axelrod, T.~S., Bloom, J.~S., Djorgovski, S.~G.,
  Fox, D.~W., Galama, T.~J., Harrison, F.~A., Hurley, K., Sari, R., Schmidt,
  B.~P., Brown, M.~J.~I., Cline, T., Frontera, F., Guidorzi, C., \& Montanari,
  E. 2002, \apjl, 572, L51

\bibitem[{Sari {et~al.}(1998)Sari, Piran, \& Narayan}]{sariEA98}
Sari, R., Piran, T., \& Narayan, R. 1998, \apjl, 497, L17

\bibitem[{Sheth {et~al.}(2003)Sheth, Frail, White, Das, Bertoldi, Walter,
  Kulkarni, \& Berger}]{shethEA03}
Sheth, K., Frail, D.~A., White, S., Das, M., Bertoldi, F., Walter, F.,
  Kulkarni, S.~R., \& Berger, E. 2003, \apjl, 595, L33

\bibitem[{Taylor {et~al.}(2004)Taylor, Frail, Berger, \& Kulkarni}]{taylorEA04}
Taylor, G.~B., Frail, D.~A., Berger, E., \& Kulkarni, S.~R. 2004, \apjl, 609,
  L1

\bibitem[{Taylor {et~al.}(2005)Taylor, Momjian, Pihlstr{\"o}m, Ghosh, \&
  Salter}]{taylorEA05}
Taylor, G.~B., Momjian, E., Pihlstr{\"o}m, Y., Ghosh, T., \& Salter, C. 2005,
  \apj, 622, 986

\bibitem[{van~der Horst {et~al.}(2008)van~der Horst, Kamble, Resmi, Wijers,
  Bhattacharya, Scheers, Rol, Strom, Kouveliotou, Oosterloo, \&
  Ishwara-Chandra}]{vanderHorstEA08}
van~der Horst, A.~J., Kamble, A., Resmi, L., Wijers, R.~A.~M.~J., Bhattacharya,
  D., Scheers, B., Rol, E., Strom, R., Kouveliotou, C., Oosterloo, T., \&
  Ishwara-Chandra, C.~H. 2008, \aap, 480, 35

\bibitem[{{van der Horst} {et~al.}(2005){van der Horst}, {Rol}, {Wijers},
  {Strom}, {Kaper}, \& {Kouveliotou}}]{vanderHorstEA05}
{van der Horst}, A.~J., {Rol}, E., {Wijers}, R.~A.~M.~J., {Strom}, R., {Kaper},
  L., \& {Kouveliotou}, C. 2005, \apj, 634, 1166

\bibitem[{Vanderspek {et~al.}(2003)Vanderspek, Crew, Doty, Villasenor,
  Monnelly, Butler, Cline, Jernigan, Levine, Martel, Morgan, Prigozhin,
  Azzibrouck, Braga, Manchanda, Pizzichini, Ricker, Atteia, Kawai, Lamb,
  Woosley, Donaghy, Suzuki, Shirasaki, Graziani, Matsuoka, Tamagawa, Torii,
  Sakamoto, Yoshida, Fenimore, Galassi, Tavenner, Nakagawa, Takahashi, Satoh,
  Urata, Boer, Olive, Dezalay, Barraud, \& Hurley}]{vanderspekEA03}
Vanderspek, R., Crew, G., Doty, J., Villasenor, J., Monnelly, G., Butler, N.,
  Cline, T., Jernigan, J.~G., Levine, A., Martel, F., Morgan, E., Prigozhin,
  G., Azzibrouck, G., Braga, J., Manchanda, R., Pizzichini, G., Ricker, G.,
  Atteia, J.-L., Kawai, N., Lamb, D., Woosley, S., Donaghy, T., Suzuki, M.,
  Shirasaki, Y., Graziani, C., Matsuoka, M., Tamagawa, T., Torii, K., Sakamoto,
  T., Yoshida, A., Fenimore, E., Galassi, M., Tavenner, T., Nakagawa, Y.,
  Takahashi, D., Satoh, R., Urata, Y., Boer, M., Olive, J.-F., Dezalay, J.-P.,
  Barraud, C., \& Hurley, K. 2003, GRB Coordinates Network, 1997, 1

\bibitem[{Waxman(1997)}]{waxman97}
Waxman, E. 1997, \apjl, 485, L5

\bibitem[{Woosley(2011)}]{woosley11}
Woosley, S.~E. 2011, ArXiv e-prints

\bibitem[{Yost {et~al.}(2003)Yost, Harrison, Sari, \& Frail}]{yostEA03}
Yost, S.~A., Harrison, F.~A., Sari, R., \& Frail, D.~A. 2003, \apj, 597, 459


\end{thebibliography}



\end{document}